\documentclass[manuscript,screen,acmsmall,authorversion,nonacm]{acmart}

\usepackage{graphicx}
\usepackage{subcaption}

\AtBeginDocument{%
  \providecommand\BibTeX{{%
    \normalfont B\kern-0.5em{\scshape i\kern-0.25em b}\kern-0.8em\TeX}}}

\setcopyright{acmlicensed}
\copyrightyear{2018}
\acmYear{2018}
\acmDOI{XXXXXXX.XXXXXXX}

\acmConference[Conference acronym 'XX]{Make sure to enter the correct
  conference title from your rights confirmation emai}{June 03--05,
  2018}{Woodstock, NY}
\acmISBN{978-1-4503-XXXX-X/18/06}

\acmSubmissionID{4377}

\begin{document}

\title[Exploring the Diversity of Music Experiences for DHH People]{Exploring the Diversity of Music Experiences for Deaf and Hard of Hearing People}

\author{Kyrie Zhixuan Zhou}
\thanks{Authors' Emails: \\ Kyrie Zhixuan Zhou: zz78@illinois.edu, Weirui Peng: wp2297@columbia.edu, Yuhan Liu: yuhanliu@sz.tsinghua.edu.cn, Rachel F. Adler: radler@illinois.edu}
\email{zz78@illinois.edu}
\affiliation{
  \institution{University of Illinois at Urbana-Champaign}
  \city{Champaign}
  \state{Illinois}
  \country{USA}
}

\author{Weirui Peng}
\email{wp2297@columbia.edu}
\affiliation{
  \institution{Columbia University}
  \city{New York}
  \state{New York}
  \country{USA}
}

\author{Yuhan Liu}
\email{yuhanliu@sz.tsinghua.edu.cn}
\affiliation{
  \institution{Tsinghua University}
  \city{Shenzhen}
  \state{Guangdong}
  \country{China}
}

\author{Rachel F. Adler}
\email{radler@illinois.edu}
\affiliation{
  \institution{University of Illinois at Urbana-Champaign}
  \city{Champaign}
  \state{Illinois}
  \country{USA}
}

\renewcommand{\shortauthors}{Kyrie Zhixuan Zhou, Weirui Peng, Yuhan Liu, and Rachel F. Adler}

\begin{abstract}
Sensory substitution or enhancement techniques have been proposed to enable deaf or hard of hearing (DHH) people to listen to and even compose music. However, little is known about how such techniques enhance DHH people's music experience. Since deafness is a spectrum -- as are DHH people's preferences and perceptions of music -- a more situated understanding of their interaction with music is needed. To understand the music experience of this population, we conducted social media analyses, both qualitatively and quantitatively, in the \textit{deaf} and \textit{hard of hearing} Reddit communities. Our content analysis revealed that DHH people leveraged sign language and visual/haptic cues to feel the music and preferred familiar, non-lyrical, instrument-heavy, and loud music. In addition, hearing aids were not customized for music, and the visual/haptic techniques developed were not widely adopted by DHH people, leading to their suboptimal music experiences. The DHH community embodied mutual support among music lovers, evidenced by active information sharing and Q\&A around music and hearing loss. We reflect on design justice for DHH people's music experience and propose practical design implications to create a more accessible music experience for them.  
\end{abstract}




\maketitle

\section{Introduction}
Music provides rich affordances, such as delicate emotional experience, enhanced connections during social events \cite{elor2021understanding}, and the potential for enhancing cognitive abilities \cite{schellenberg2005music}. The main barriers to an optimal music experience for deaf or hard of hearing (DHH) people are their limited access to the audio features of music \cite{silva2020music}. Children with cochlear implants (CIs) showed the ability to perceive emotions (e.g., happy vs sad) in music but did so less accurately than typically hearing peers \cite{hopyan2011identifying}. 
Interviews with musicians with hearing loss revealed a low satisfaction with digital hearing aids (HAs) -- pitch and timbre distortion was often reported to compromise music listening experiences \cite{fulford2015hearing}.

Music and music education should be made more accessible for DHH people \cite{wright2014bridging, kyriakou2022teaching}, who can benefit from music -- music has been shown to improve the cognitive performance of deaf children \cite{rochette2014music} and help them acquire new vocabularies through music therapy \cite{bassiouny2017using}. Accessibility for DHH people is also an important consideration in live music events such as music festivals \cite{bossey2020accessibility}. DHH people expressed difficulty joining sound-based communal activities such as fan chanting and suggested utilizing visual stimuli such as AR glasses to provide contextual information \cite{han2023opportunities}. Assistive music technologies, mostly drawing on haptic \cite{fahey1972education, cavdir2020felt, jack2015designing} and visual \cite{deja2020vitune, grierson2011making, kim2015seen} cues have been designed for DHH people. However, how DHH people experience music, potentially with the help of such assistive technologies, is not well understood. 

Toward bridging this research gap, we analyzed 535 music-related posts/comments in the r/deaf and r/hardofhearing communities on Reddit. The haptic and visual music technologies developed in academia were not widely adopted in our sample. The Reddit users mostly drew on their own experiences and skills, as well as external help from the community, to optimize their music experience, e.g., drawing on sign language and visual/vibrational cues, listening to familiar and specific types of music, and sharing information about accessible music and devices. They still experienced a wide range of accessibility challenges with music such as mishearing lyrics and a lack of captions in online videos and movies. After presenting a situated understanding of the experiences and challenges of DHH people in music through social media analyses, we provide suggestions for contextual, affordable, technology-assisted designs to create a more accessible music experience.

\section{Related Work: Assistive Music Technologies for DHH People}
DHH people have limited access to auditory information, thus they need to rely on other senses to feel the music. Researchers have 
called for multimodal music streaming experiences for DHH people \cite{mchugh2021towards}. Visual and vibratory stimuli are
the most cited methods of experiencing music by DHH people themselves \cite{watkins2017deaf}, sometimes used together (e.g., \cite{petry2016muss}). Below, we elaborate on assistive music technologies that leveraged these two senses.

\subsection{Visual Music} 

DHH people use assistive technologies such as sound visualization and speech-to-text to improve sound perception via a combination of auditory and visual means \cite{ohshiro2022people}. Fourney and Fels explored music visualization to help hard of hearing (HoH) music consumers understand the emotions conveyed by music \cite{fourney2009creating} -- they used a piano roll-up to represent notes and a moving ball to represent note lengths. 
Deja et al. presented an on-screen visualizer that generated effects alongside music -- it was shown to effectively enhance the musical experiences of DHH people \cite{deja2020vitune}. Grierson developed an interactive audio-visual performance system for deaf children's music interaction \cite{grierson2011making}. Kim et al. demonstrated tangible music visualizations that leveraged items children saw in their daily lives, such as flowers and plants \cite{kim2015seen}.

Lyrics were perceived by d/Deaf people as one of the central elements in a song -- song signing augmented their musical experiences with an additional level of accessibility \cite{yoo2023understanding}. Maler argued that ``d/Deaf song signers embody music differently than the hearing by creating a visual, kinetic form of music in sign language, rather than using sign language to express something about sound'' --  deafness may thus become a source of musical ability rather than a source of impairment \cite{maler2015musical}.

\subsection{Haptic Music} 

Perceiving music by touch has been extensively investigated \cite{fahey1972education, cavdir2020felt, jack2015designing, jackdesign, tranchant2017feeling, araujo2016auris, yan2022adaptive, chao2018dancevibe, remache2021audio, aker2022effect, baijal2012composing}, with pitch, rhythm, loudness, and timbre mapped to tactile vibrations. Displaying emotionally resonant sounds as vibrotactile feedback could produce distinct emotional responses \cite{sion2023me}.

Rhythm is often the first musical concept DHH people learn in music classes. Florian et al. developed a prototype to generate light and vibrotactile outputs from music rhythm for deaf people \cite{florian2017deaf}. Providing ad-hoc, real-time rhythm information to a deaf user was deemed important \cite{petry2016ad}. Deaf people were even able to collaborate with hearing people to create short beats with the help of sound visualization and haptic feedback \cite{soderberg2016music}.

Haptic sense is also helpful in delivering music as a whole to DHH people. Karam et al. presented a sensory substitution technique that delivered music as discrete channels of vibrotactile stimuli, perceived positively by Deaf senior citizens \cite{karam2009modelling}. Nanayakkara et al. combined a ``Haptic Chair'' and a computer display of visual effects to enhance DHH people's music experience \cite{nanayakkara2013enhancing, nanayakkara2009enhanced}. Trivedi et al. aimed to develop an affordable wearable haptic device, which turned out to be vibrotactile sleeves with bone conduction speakers \cite{trivedi2019wearable}. Audio-tactile music interactions were further integrated into immersive VR music videos \cite{young2023feel}. Deaf children felt more confident when wearing a visual and vibrotactile music-sensory substitution device \cite{petry2018supporting}. Musical vibrations also played an important role in music therapy with D/deaf clients \cite{palmer2022vibrational}. 

Further, accessible Digital Musical Instruments (ADMIs) have been extensively explored in inclusive music practice, with an emphasis on tangible or physical controllers \cite{frid2019accessible}. Iijima et al. prototyped a ``smartphone drum,'' a smartphone application that presented a drum-like vibrotactile sensation when users used smartphones as drumsticks and made drumming motions in the air  \cite{iijima2021smartphone}. End users’ preferred gestures when using smartphones to simulate instrumental experience were identified \cite{iijima2022designing}.

\subsection{Research Gap} Despite the fruitful designs of assistive music technologies, as shown above, they were not widely adopted by DHH people -- their expressed barriers to technology included unknown availability and limited options \cite{ohshiro2022people}. Limited research has been conducted to understand how DHH people practically experience music and leverage assistive technologies (e.g., hearing aids, headphones, advanced haptic/visual devices) to enhance their music experience. In this paper, we explore the diverse, rich discussion in DHH communities on Reddit to add to this literature.

\section{Methodology}

\subsection{Data Collection}
We searched both the r/deaf and r/hardofhearing subreddits in October 2023 using the keyword ``music.'' We examined the top 40 posts (by relevance), 20 in each subreddit, ranging from 8 years old to 1 month old by the time of the search. 
The posts concerned general music experience (e.g., ``How do you feel about music?''), music recommendations for DHH people (e.g., ``Music for deaf kids?''), device recommendations (e.g., ``Deaf on left ear looking for BT in-ear headphone for iPhone and music''), ways to better experience music (e.g., ``How do you feel music through vibrations?''), accessibility challenges or features (e.g., ``App that adds subtitles to many apps''), and medical questions (e.g., ``hearing loss from loud music?''). Comments under each post were collected, leading to 352 and 183 comments, respectively.  

\subsection{Quantitative Analysis}
To understand the sentiments and themes prevalent in the DHH communities' discussions about music, we applied topic modeling and sentiment analysis to the collected data. Comments that were off-topic, contained spam, or lacked substantive content were not constructive to the analysis. To ensure the relevance and quality of the quantitative analysis, we implemented a filtering process. We evaluated the comments' relevance to the DHH community's interaction with music. Meaningless or low-effort posts, such as those with minimal textual content or unrelated to DHH people's music experience, were filtered out by the authors. After this filtering process, our final dataset for the quantitative analysis comprised 291 posts. 

\subsubsection{Sentiment Analysis}
Sentiment analysis was conducted using TextBlob, a library offering a straightforward API for natural language processing tasks \cite{diyasa2021twitter}. TextBlob computes sentiment scores by evaluating the polarity of words in the text, with scores ranging from -1 (highly negative) to +1 (highly positive). These scores facilitated the classification of sentiments into three categories: negative (scores less than -0.01), neutral (scores between -0.01 and 0.01), and positive (scores greater than +0.01).

\begin{itemize}
    \item \textbf{Negative Sentiment:} Text with scores within $[-0.1, -0.01)$, reflecting expressions of dissatisfaction or concern. Examples are shown below,
    \begin{quote}
        ``I don't listen to songs, I don't have even one song on my iPhone, I listen to other people's songs probably once or twice a year. I roll my eyes at those who sign songs on YouTube (though there's a gem sometimes). I sometimes find music annoying, actually. It's just nonexistent in my life. The same cannot be said for other Deaf people -- I've seen a lot of Gallaudet students listening to music, they probably have milder hearing loss and can relate to music, whereas I am profoundly deaf and I see no point in music.''
    \end{quote}
    \begin{quote}
       ``My dad always blasted the radio. I don't fully understand how this is marked 'deaf/hoh with questions' if you don't understand how this works?''
    \end{quote}
    
    \item \textbf{Neutral Sentiment:} Text with scores within $[-0.01, +0.01]$, denoting a balanced or indifferent emotional tone. Examples are shown below,
    \begin{quote}
        ``HoH here. I dont hear bass and I cant make out lyrics. Recorded music i use bone conductors, and live music i listen by feeling vibrations. If i am in a restaurant or bar or something with music in the background, i probably cant make any of it out, if i detect it at all''
    \end{quote}
    \begin{quote}
       ``I'm profoundly deaf, however, I like to turn on the volume to its max and blast my music. I love feeling the bass. I prefer to use music videos from YouTube so I don't look like an idiot blasting radio. Funny story - when I was a teenager riding in a deaf friend's car blasting `music,' and pulled up to my hearing parents, they came out and said we were listening to Oprah's Talk Show. I like dance/house because it has such a strong bass.''
    \end{quote}
    \item \textbf{Positive Sentiment:} Text with scores within $(+0.01, +0.1]$, indicating a positive or affirmative disposition. Examples are shown below,
    \begin{quote}
        ``Damn thats awesome! I have no info about it but its really cool and nice and i hope you find help for this :)''
    \end{quote}
    \begin{quote}
       ``I listen to music quite a lot, in different genres and have done for most of my life. It makes me feel a variety of emotions like hearing people.''
    \end{quote}
\end{itemize}

To visually assess the sentiment distribution, we utilized histograms and count plots. Histograms showcased the overall sentiment tendency, while count plots provided a frequency comparison of the sentiment categories. 

\subsubsection{Topic Modeling}
We began by pre-processing the text data from our dataset, including tokenization, normalization, and lemmatization using NLTK's WordNetLemmatizer. This process also involved the removal of stopwords to enhance the relevance of our feature set. We then vectorized the processed content using TfidfVectorizer, which transformed the text into a TF-IDF matrix -- a numerical representation that reflected the importance of words within the corpus. K-Means clustering was chosen for its effectiveness in partitioning the words into ten distinct clusters, each potentially representing a topic. By fitting the TF-IDF matrix into the K-Means model, we identified centroids for each cluster, which were indicative of the central themes of each topic. To interpret these clusters, we extracted the top ten keywords for each, using the centroids to locate the most defining words within our TF-IDF feature space. These keywords served as a proxy for the topics discussed within the community.

\subsection{Qualitative Analysis}
After uncovering the nuanced and diverse nature of DHH communities' discussion about music, music technology, and hearing impairments, we further employed a collaborative thematic analysis \cite{braun2012thematic} to gain more in-depth insights into the posts and comments. Two researchers performed the analysis independently, generating themes and subthemes, and regularly discussed to reach a consensus. The three overarching themes we summarized were ``Understanding DHH People's Music Experience'' (Section \ref{understand}), ``How DHH People Enhance Their Music Experience'' (Section \ref{how}), and ``Information Sharing Within DHH Communities'' (Section \ref{sharing}). Within these themes, subthemes emerged. For example, under ``How DHH People Enhance Their Music Experience,'' subthemes included ``familiar music,'' ``non-lyrical, instrument-heavy, and loud music,'' ``ASL (American Sign Language),'' ``visual music,'' and ``haptic music.'' A theoretical saturation was reached after we coded 10 posts in each community, i.e., no new themes emerged. We used XMind, a mind-mapping tool, to organize different levels of themes and quotes into a hierarchical structure. In the paper, we use raw quotes from Reddit users to illustrate our findings. Notably, HoH people tended to refer to their hearing impairments as ``hearing loss.'' Otherwise, the differences in the discussions between the deaf and HoH communities were not obvious. Unless explicitly differentiated, our findings apply to DHH people in general. 

\section{Findings}

\subsection{Quantitative Results}

\subsubsection{Sentiment Analysis Results}
\begin{figure}
  \centering
  \begin{subfigure}{0.45\textwidth}
    \includegraphics[width=\linewidth]{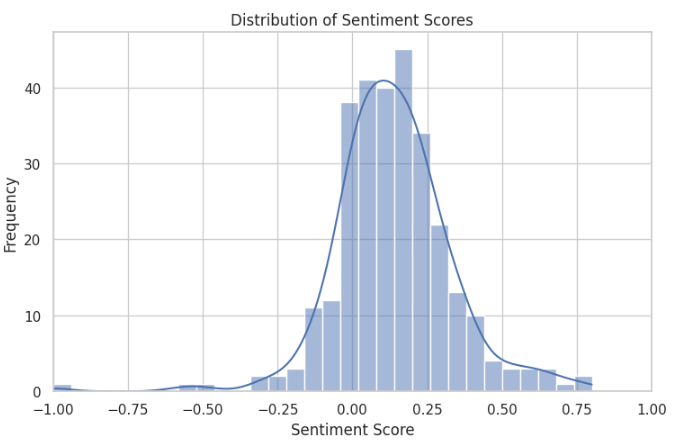}
    \caption{Histogram of sentiment scores.}
    \label{fig:histogram}
  \end{subfigure}
  \hfill
  \begin{subfigure}{0.45\textwidth}
    \includegraphics[width=\linewidth]{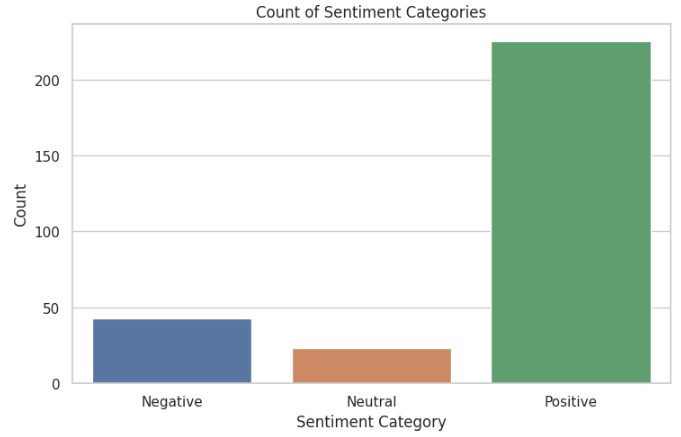}
    \caption{Count plot for different sentiments.}
    \label{fig:count}
  \end{subfigure}
  \caption{Visualization for sentiment analysis results.}
\end{figure}

We utilized a histogram to demonstrate the distribution of sentiment scores, as shown in Figure \ref{fig:histogram}. The analysis uncovered a slight right-skewness in the sentiment distribution, with the bulk of sentiment scores clustering around the neutral to slightly positive range. This skewness towards a more positive sentiment indicated a generally favorable dialogue within the DHH community regarding their music experiences and their use of assistive technologies. Despite the potential challenges that the DHH community may face, the prevailing sentiment captured in the dataset suggested a predominance of positive expressions, which could be reflective of positive personal experiences, supportive initiatives, or progress in technology that enhanced music accessibility.

Moreover, as depicted in Figure \ref{fig:count}, we categorized the sentiment of the posts and comments into three distinct groups: negative, neutral, and positive. The count plot illustrated a predominant number of texts with positive sentiment (N=225), significantly outnumbering those classified as neutral (N=23) or negative (N=43). This distribution again suggested that the majority of the community's discourse was optimistic or constructive when discussing music experiences. A smaller, yet notable number of texts were categorized as neutral, indicating a balanced or impartial view. Meanwhile, the posts with negative sentiment represented a less common sentiment category, which may point to the few instances of dissatisfaction or adverse experiences within the dataset. This visual representation reinforced the notion that, while challenges existed, the overall sentiment within the community conversations was largely positive.


The sentiment analysis underscored the complexity of the emotional responses within the DHH community. While the histogram in Figure \ref{fig:histogram} revealed a slight right-skewness in sentiment scores, indicating a tendency towards positivity, the count plot in Figure \ref{fig:count} further corroborated this finding by showing a predominant number of posts with positive sentiment. These optimistic expressions significantly outnumbered the neutral and negative categories, suggesting that the majority of the community's discourse was constructive when discussing music experiences and the use of assistive technologies. Despite this, the presence of negative sentiment, albeit as a less common sentiment category, highlighted that certain aspects of the DHH community's experiences did warrant attention. 

\subsubsection{Topic Modeling Results}
We employed a K-Means clustering approach on the TF-IDF vectorized content of the DHH community's Reddit discussions to discern underlying themes. 
Ten clusters were identified, with each cluster representing a distinct topic prevalent within the DHH communities' discussion of music.

The top keywords from each cluster offered a glimpse into the thematic focus areas, as shown in Table \ref{tab:topic_modeling}. For example, Cluster 0 highlighted discussions centered around hope, understanding of the brain's interaction with music, and general knowledge about deafness and hearing; contrastingly, Cluster 1 was characterized by practical discussions about hearing devices such as headphone and AirPods. In clusters focusing on ``Music Experience'' (Cluster 3 and Cluster 7), terms such as ``enjoy,'' ``loud,'' ``song,'' ``lyric,'' and ``bass'' indicated a rich dialogue on personal interactions with music, highlighting both the sensory aspects of music and the emotional connections to music. Clusters like Cluster 2 and Cluster 6, which were enriched with keywords like ``hearing,'' ``aid,'' ``loss,'' ``audiologist,'' and '`Bluetooth,'' reflected a technological slant, discussing the use of hearing aids and other assistive technologies, as well as the implications of hearing loss. Cluster 4 and Cluster 8 delved into the emotional and social dimensions, with words such as ``care,'' ``appreciate,'' ``health,'' ``deaf,'' ``people,'' ``love,'' ``friend,'' and ``story.'' These clusters seemed to capture the community's sentiment towards hearing health and the social aspects of being DHH, including the shared stories and relationships that formed around music. The remaining clusters, such as Cluster 5 and Cluster 9, contained terms that suggested a focus on practical solutions and the sensory experience of music, respectively. Words like ``mono,'' ``sound,'' ``channel,'' ``feel,'' and ``loss'' seemed to point towards a dialogue on the technical adaptations for mono hearing and the tactile experience of music.


Overall, the K-Means clustering and the examination of the top keywords in each cluster provided a comprehensive thematic map of the discussions within the DHH communities on Reddit. 

\begin{table}[H]
\centering
\caption{Top 10 keywords in each cluster from topic modeling.}
\label{tab:topic_modeling}
\begin{tabular}{|c|p{11cm}|}
\hline
\textbf{Cluster} & \textbf{Top 10 Keywords} \\
\hline
0 & would, brain, music, hope, think, know, deaf, hear, find, hearing \\
\hline
1 & headphone, call, airpods, pretty, charge, clarity, well, single, might, work \\
\hline
2 & hearing, aid, music, bluetooth, sound, loss, without, ear, car, use \\
\hline
3 & music, hoh, im, still, always, enjoy, loud, thing, loved, lyric \\
\hline
4 & know, hearing, dont, care, sometimes, appreciate, music, better, health, loss \\
\hline
5 & left, right, check, solution, bit, mono, sound, perhaps, combined, channel \\
\hline
6 & time, get, audiologist, better, make, noise, bad, going, one, good \\
\hline
7 & song, hear, bass, lyric, like, music, playing, instrument, play, im \\
\hline
8 & music, deaf, listen, people, dont, love, friend, story, party, find \\
\hline
9 & music, sound, like, im, hearing, ear, feel, would, help, loss \\
\hline
\end{tabular}
\end{table}

\subsubsection{Interpretation}
The combined sentiment and topic analyses provided a concise yet comprehensive portrayal of the DHH community's online discourse around music. Sentiment scores varied yet showed an inclination towards positive emotions when discussing music experiences. The topic modeling unearthed a nuanced facet of the community -- a collective that not only confronted difficulties but also celebrated the joys of music and music technologies. The presence of both positive and negative sentiments underscored the complexity of the DHH experience and highlighted essential considerations for the design of assistive technologies. The results from the sentiment analysis and topic modeling painted a picture of a community actively navigating the intersection of music, technology, and hearing impairments. 



\subsection{Qualitative Results}
\subsubsection{Understanding DHH People's Music Experience}
\label{understand}
Hearing people may have biased assumptions about DHH people's music experiences, which annoyed them constantly in the Reddit discussions. It is important to learn and understand DHH people's music experience instead of imposing assumptions. Both DHH people's diverse music experiences and their accessibility challenges with music are discussed below.

\noindent \textbf{DHH Individuals' Diverse Music Experiences.} The DHH community was annoyed when people without hearing impairments held biased assumptions about DHH people's experiences, preferences, or attitudes toward music. Assuming every DHH person liked music by default was rude and regarded as a hearing-centric perspective. For example, when a hearing user asked what album to buy for their deaf niece as a gift, a deaf user argued that not every deaf person would eventually listen to music, and it was nicer to understand deaf people's world, instead of trying to integrate them into the world of people without hearing loss, \textit{``You have an opportunity to enter into a new understanding of a world and a culture that is hers by rights. It’s a rare and truly beautiful thing if you let it be. Rather than rushing to integrate her into your world (one in which she will always be an outsider), strive to first learn hers.''} When a Reddit user asked if deaf people listened to music, a member of the deaf community argued that music experiences and preferences varied among people and that deaf people should not be treated as one, \textit{``We are not a monolith. You should just ask them! I for one don't enjoy music at all. I know there are deaf people out there [who enjoy music], and one in particular has a YouTube channel. You should find an actual deaf person and ask them what a deaf space might look like.''} Similar annoyance occurred when hearing people assumed all DHH people had the same music preferences, \textit{``I kinda get what you are asking, but would ever pose the question `what sort of music do POC [people of color] like? Or what kinds of music do Jewish people like?' Can you see how it reduces us to the sun of all people, rather than letting us feel like individuals which is a big part of the reason we are in these spaces?''}

Deafness is a spectrum, which may lead to different music experiences. One deaf user elaborated on this opinion, \textit{``It's important to note that deafness is a spectrum, and not everyone who is deaf has no hearing. Everyone has different levels of how much and what they can hear... I have heard of profoundly deaf people who enjoy the vibrations of music. I am deaf or hard of hearing, and I can hear music as long as it's loud enough.''} Another user echoed this view, \textit{``There’s a huge range of hearing levels -- ranging from a mild hearing loss to hearing nothing. Deaf (and hard-of-hearing) ppl are not a monolithic group. We are regular individuals with different tastes and preferences.''} Compared to deaf people, HoH people tended to enjoy music more. 
Some of them even became professional musicians, with help as simple as good headphones, \textit{``I'm an active musician despite my hearing loss! I've released a few albums and I play bass. I find that with technology I don't have much of a problem; I just have to keep the volume of things up pretty high and have good headphones. And learning about things like waveforms so I can visually identify any issues too :)''} 

DHH people's preferences for music can be contextual. For example, music was not wanted at parties, where people needed to talk, \textit{``Basically, listening to what people are saying takes a lot of energy but it doesn't discriminate easily if a radio is playing in the background. I can't just tune it out, it's just as present as what someone is saying.''} Several DHH people described listening as work for them. Having music as the background made listening and conversing much more difficult.
When a hearing user planned to hold a party for deaf people and asked about music choices, a deaf user explained how music made the party ``deaf-UNfriendly'': \textit{``
If we wanted to use speech to text to talk with someone, the music would interfere. If we wear assistive listening devices, the music will be distorted, staticky, etc. You came at this from the wrong direction. You’re having a party, not a concert, and party suggests socializing more than enjoying music. You have to make the socializing aspect deaf-friendly, not the music taste.''}

Further, some DHH people preferred to enjoy music alone instead of treating it as a group activity, given their extra challenge with music,
\textit{``I seldom enjoy music with others. I'm okay with bopping along to a familiar song, but I loathe feeling like I have to perform. It's likely because as a kid, I was forced to participate in music class when I knew I couldn't follow along and sang gibberish. If I feel motivated to pretend I'm Slash with the air guitar, then I will, but if not, I'm not budging. Music is not supposed to be a chore.''}

\noindent \textbf{Accessibility Challenges with Music.} Hearing loss was a barrier to effectively listening to or enjoying music. Some DHH people gave up music which became random noise to them. This user was an example: \textit{``After losing my hearing and getting a cochlear implant, music has become dull, mechanical, and most of all, indistinguishable from a wall of aluminum cans crashing to the floor. None of it holds any meaning to me anymore. So, I don't bother with music anymore.''} Those who were profoundly deaf may find even more difficulty listening to music,  \textit{``I don't listen to songs. I don't have even one song on my iPhone. I listen to other people's songs probably once or twice a year. I roll my eyes at these who sign songs on YouTube (though there's a gem sometimes). I sometimes find music annoying, actually. It's just nonexistent in my life.''} DHH people who did not enjoy music may nevertheless use music for other personal purposes, e.g., to block noise from colleagues at work, \textit{``I don't give a shit [about music]. While I think some songs can be lit af, I can't pick out one particular song by sound alone. I do listen to it a lot at work, but that's to block out the people laughing too loudly near my cubicle.''}

Accessibility infrastructure did not always exist to help DHH people listen to music. Some people complained that contextual caption was often missing in movies or online videos when music was played as the background or as a hint for the atmosphere. Without contextual, visual (i.e., textual) descriptions of music, DHH people may find difficulty understanding the messages conveyed by music, especially if the music was not loud, \textit{``Just [music playing] is lazy. Especially since most of the time doesn't differentiate between music in the story and out of the story and doesn't tell you what kind of music is being played... If a character enters a house and hears piano, then it's valuable information. There's probably someone in the living room playing piano. If it's loud music then you get why the character is annoyed. Even atmospheric music is useful if the kind is given.''} An HoH user talked about interpretation challenges brought by missing captions for YouTube and TikTok videos, \textit{``If a YouTube video has no captions, I usually can’t watch it :( I also just don’t use TikTok because it’s so hard to understand most of the videos.''} 

Even slightly more detailed captions for music might be of help, \textit{``Well, I like `ominous music plays' better than `music playing'''} Netflix was mentioned as a good example of providing contextual captions of music, 
\textit{``If the song has any importance to the plot, I think descriptions are awesome. Grey's Anatomy is especially good about that, they even make sure to put the title of the song, the artist, and the lyrics. But if the song is only there for background music, a simple `Soft music starts' is fine if it seems like it is anticipating something, or just omitted if it's just there to fill in silence.''} 

Several HoH users recommended the live captioning function in iOS and Google Live Transcribe. One of them raved about the accessibility features provided by Google Pixel and Apple, \textit{``The Pixel were the very first to come out with Live Captioning, Live Transcribe, Sound Notifications (let you know if a baby is crying or an alarm is going off), Call Hold (Alert you when the call hold is done), and Direct My Call (listen to the automated voice menu and lists the options for you). All of these features are on the Pixels first and then sent out to other Android phones in later OS updates. Apple will copy them and improve on the features. Win/win for all of us! I think the Pixels are the best for the Deaf/HoH who use their voice to communicate with hearing people. A huge kudos for Apple to caption FaceTime and Google to caption Duo \& Meet!''} Another HoH person shared how Airpods compensated for their hearing capability, \textit{``I have around 40dB loss, 60/70 as you get towards the higher frequencies, and I find that the airpods are fantastic at filling in the gaps, and the pros make phone calls way easier.''} More advanced smartphones tended to provide better captioning performance, \textit{``For Android users, I'll recommend getting the Google Pixel 6 or higher. The new Tensor chip really pulls its weight in providing a more accurate captioning even if the phone is in airplane mode.''} This might impose additional challenges on socioeconomically disadvantaged smartphone users when they try to access captions for music. 

\subsubsection{How DHH People Enhance Their Music Experience}
\label{how}
DHH people in our sample did not extensively use music technologies developed in academia but used their own heuristics to listen to music, such as sticking to familiar music, actively choosing non-lyrical, instrumental-heavy, and loud music, relying on ASL, and leveraging visual and haptic cues. 

\noindent \textbf{Familiar Music.} DHH people often had to listen to music multiple times to pick up the tune, \textit{``I have to listen to something multiple times before I start to pick up the tune. So live music still sucks.''} Due to the difficulty of picking up new music, they tended to listen to music that they were familiar with, \textit{``If I have heard the song before and know the lyrics and the music, I can hear it better.''} Even a semi-pro musician shared this habit, \textit{``I'm HoH (moderate to severe). I have my bachelor's degree in music theory and am a semi-pro musician. However, I have a complicated relationship with music. People are often surprised, given my involvement with and knowledge of music, that I don't listen to that much music... If I do have music on in the background, it's almost always music I've heard many many times and is extremely familiar.''}

\noindent \textbf{Non-lyrical, Instrument-heavy, and Loud Music Is More Accessible.} Instrument-heavy and loud music was favored by most DHH people, and deaf people in particular liked non-lyrical music. 

Recognizing words was challenging for deaf people. Thus they preferred non-lyrical, non-vocal music such as electronica, \textit{``Well with hearing aids most of the time you could hear it like anyone else but to me as near deaf I can still hear the beat and I enjoy the music words not very much and maybe instrument of loud enough but for me I just hear the beat and honestly like that.''} Another deaf person also expressed that lyrics were intelligible and they preferred instrumental types of music, \textit{``
In my case, I can hear most instruments fairly clearly, but the lyrics of course are unintelligible. I tend to prefer instrumental types of music for that reason - mostly chiptunes/synthwave, but also some jazz or lofi hiphop at times.''}

Instrument-heavy music could be better felt by DHH people, e.g., \textit{``I tend to prefer bass-heavy music since I can actually feel it and move along when in a club or party setting.''} \textit{``I find that music with strong percussion sounds best in my CI.''} \textit{``Rap music will have good bass. What are you planning on using for streaming it? Increase the bass and it will vibrate with most songs.''} One person preferred instrumental music and noted that they were only able to fully enjoy lyrical music if the lyrics were provided, \textit{``I listen to mainly instrumental music myself. Have to listen in a quiet room since my `ears' will pick up outside sounds... I do like some pop/country/bluegrass but need the lyrics in front of me to fully enjoy.''} Another person directly asked for suggestions for music with bass, \textit{``Hey, my friend has some music clips that plays when chilling with his friends. I was thinking, is there some music that's just loaded with good rhythmic bass? Sorry if this is a dumb question, and thanks if you could share a playlist or some music.''} 

Loud music was more comprehensible, thus DHH people tended to ``blast music'' or ``blast speakers/radios.'' One user described their preferred level of loudness, \textit{``This is individual, of course. But generally loud. When I was at Gallaudet, the parties in a specific dorm were so loud hearing students often would not live there. The walls and furniture would shake from the sound.''} Some played music loud to feel the bass, \textit{``I'm profoundly deaf, however, I like to turn on the volume to its max and blast my music. I love feeling the bass.''}
Loudness helped HoH people better understand lyrics, \textit{``The more vibrations from the music the better if I cannot make out the lyrics.''} Another HoH person attributed the effect of loudness to the association between vibrations and sound in memory, \textit{``What I do now is annoy anyone around me by insisting on turning up the bass as high as I can because although I can’t hear it, I can feel it and it is nearly the same to me now. I feel like I’m hearing it because the vibrations take the place where the sound once would have been in my brain, is my theory, but most people tell me it ruins the song and vocals. To me, it makes it sound like it used to sound!''} Even those who did not like loud music acknowledged that easy-listening music made subtle changes in pith undistinguishable, \textit{``I seem to have no tolerance for loud fast songs with a lot of clutter and noise. Most of the stuff I listen to is easy listening. But it can't be too easy because I don't distinguish between more subtle changes in pitch...''}

Playing music loud often annoyed family members or neighbors. The following example illustrated how a HoH photographer negotiated between having superior music experience and not bothering her husband, \textit{``I almost only get to listen when I drive and then I get to have it at a level that I can hear it. My husband says it's too loud for him. I have one speaker and it sits on a part of my desk about ten inches from my hard-of-hearing side. I always ask if it's too loud, because our dining room (where my desk is) and the living room are open to each other. I keep my music quiet enough to not drown out his TV and he keeps the TV low enough that it doesn't drown out my music.''} The ``rude loud deaf neighbor'' discussion often occurred. DHH people agreed that it was okay to request a volume adjustment when one's deaf neighbor played music too loud. They also suggested headphones that reserved optimal music experience while not annoying neighbors. Some showcased how they dealt with this dilemma, e.g., reaching a consensus on socially acceptable ``loud hours'' with neighbors, \textit{``Completely deaf here. I have some very loud hobbies like DIY, metal bashing, cutting up wood with power saws for my carpentry/repair hobbies, drilling holes in walls, etc. I ALWAYS check with my neighbors over what they consider socially acceptable hours for noisy hobbies and try to stick to them.''}

The Reddit users enthusiastically shared songs, singers/bands, albums, and Spotify playlists that were accessible to deaf people, often with the above-mentioned features, e.g., \textit{``Most songs by RHCP since Flea is an absolute legend on the bass.''}

\noindent \textbf{ASL.} DHH people tended to use sign language to interact with music and each other, \textit{``I often will tell people that ASL is loud...visually. When there is a party or an event with a large number of Deaf people, I don't `miss out' on music because there is music playing with people's hands.''} Some even hired sign language interpreters to sign music for them at home. 

Musical venues were obligated to provide interpreters for DHH people. When a deaf user asked how they could enjoy music at a rock show, another user suggested that the venue was responsible for finding an interpreter for them. The National Association of the Deaf (NAD) had an advocacy letter to ask stubborn venues to comply: \textit{``If you're in the US, the responsibility of obtaining an interpreter rests entirely on the venue and they cannot charge you extra for your request. The NAD also has an advocacy letter\footnote{\url{https://www.nad.org/wp-content/uploads/2020/06/Festivals-and-Concerts.pdf}} for more stubborn venues that try to push back that clearly explains the relevant laws they need to follow. I have a lot of experience with these sorts of requests and the pushback venues will give to us.''} Someone else also mentioned that providing qualified interpreters for deaf people was a federal law, \textit{``This! And they have to give you a QUALIFIED interpreter. You can’t have your kids/spouse/friend just interpret for you. It’s federal law. I know it’s scary but you’re in the right here. You’ve got this :)''} Dropping the law bomb was often seen as a last resort when DHH people went to musical venues. 

\noindent \textbf{Visual Music.} A lot of DHH users on Reddit mentioned the combination of audio music with visual elements, such as colors and lighting: \textit{``I feel like HOH or deaf people are more focused on visual cues since one of your senses is not working or working less. Also I’ve experienced and got to know that HOH/dead people are often way more sensible to things such as light.''} \textit{``I love when music is paired with lighting personally, or strong bass''} One HoH user used visual features to enhance their viola playing experience, \textit{``I have bilateral cochlear implants and take viola lessons. I think keeping time involves developing an inner pulse, and learning to use a metronome - especially one with visual feature, is one way to develop that inner pulse.''}

In the HoH community, people commonly expressed challenges with lyrics, e.g., mishearing lyrics, \textit{``Mishearing song lyrics is a common thing, even for those who don't have hearing loss. Of course, it's worse for us. There are even websites devoted to misheard lyrics. Some of the misheard lyrics are hysterical.''} For many, the challenges did not disappear even when using hearing aids or cochlear implants. One user gave a vivid example of how lyrics could be misheard, \textit{``*Cheese is on the shelf in there.* Then I finally looked at the satellite radio display which was playing the Elvin Bishop song `Fooled Around And Fell In Love'.''} Many people referred to such mishearing experiences as hysterical or hilarious. 
Videos showing lyrics with music in the background enabled them to interpret lyrics, although they could not be watched anywhere, anytime, \textit{``I'm hard of hearing so with my hearing aids I can hear mostly everything. With them I just enjoy the beats but can't understand most of the lyrics. But I read the lyrics on YouTube lyrics videos a lot while I'm listening to songs. Obviously I can't watch the lyrics videos while driving tho.''} 

Other favorable visual elements included captions and body language. Some people even solely relied on visuals of music, \textit{``I find Deaf musicians and watch (not listen) their work.''} A Reddit user shared a music video created by a deaf artist, who told the story of social isolation and the ultimate ``coming home'' to the shared Deaf/HoH community, \textit{``It is a testament to our prevailing power and the strength of community.''}

\noindent \textbf{Haptic Music.} Haptic/vibrational/tactile music was mentioned by multiple people. One HoH user listened to music through vibrations: \textit{``HoH here. I don't hear bass and I can't make out lyrics. Recorded music I use bone conductors, and live music I listen by feeling vibrations. If I am in a restaurant or bar or something with music in the background, I probably can't make any of it out, if I detect it at all.''} Some were amazed by how others could sing by looking at the mouth and feeling vibrations, \textit{``I want to know how to enjoy music after losing my hearing, as a person who once enjoyed it a lot. A curious part I know, there are a few people who actually learn how to speak and sing looking at the mouth and feeling vibrations equivalent to some notes.''}

Several academic or industrial projects on haptic music technologies for DHH people were mentioned and linked\footnote{\url{https://www.iq-mag.net/2020/08/vibrating-vest-hear-music-through-skin/}}\footnote{\url{https://www.classicfm.com/discover-music/live-music-deaf-audiences-vibrating-vests/}}\footnote{\url{https://www.wired.co.uk/article/understanding-the-brain-david-eagleman}}. One user elaborated on a research project conducted by the University of Liverpool: \textit{``Your friend may be interested to know that the University of Liverpool is undertaking research\footnote{\url{https://www.musicalvibrations.com/}} into using Vibrotactile technology to support deaf people in music education, performance, appreciation and production.''} An HoH person talked about their satisfactory experience of using a vibrotextile wearable to listen to music at a musical, \textit{``I'm HoH and was at a small free event last year put on by Opera Philadelphia that had Vibrotextile wearables from Music: Not Impossible. It was pretty cool.. a vest and a strap on each arm which had many different vibration patterns to experience the sound. I'm not sure where they're at as far as lending out for use at events but they're @musicnotimpossible on Instagram and if you Google `Music Not Impossible' you'll find their site.''} However, such devices have not been produced at scale and are not accessible to the DHH community, as evidenced in the Reddit discussions. 

Without advanced devices, the music experience created by vibration could hardly be optimal, \textit{``No, I can’t clearly comprehend music through vibrations of speakers. If I want to understand a song, I put on my hearing aid or turn up the volume until I can hear it. Vibrations are just the bass, beat, tempo, general sound, or whatever the musical terms are (I’m an idiot when it comes to music).''} Higher frequencies were harder to feel via vibrations, \textit{``Higher frequencies need to be several orders of magnitude louder/stronger for them to be felt. So while a person can get more from touching the speaker it's nothing like hearing the music, but if you've been deaf your whole life, you wouldn't know the difference.''}

\subsubsection{Information Sharing Within DHH Communities}
\label{sharing}
Information sharing and mental support regarding both music and hearing loss are common in Reddit communities.

\noindent \textbf{Information Sharing for Better Music Experience.} DHH people frankly shared their suboptimal music experience even with hearing aids or cochlear implants: \textit{``Does it help? Yeah, it satisfies my need. Do I hear as well as a hearing person? Nah, definitely not, but I don’t need to hear all the little nuances to enjoy music. If I want to understand lyrics, I read the lyrics (I can’t understand words without seeing them first). I can’t tell instruments apart from each other, but that’s fine. I don’t need to know that.''} One HoH person noted that hearing aids were not tailored for music experience, \textit{``The main program in hearing aids is adapted to increase speech comprehension, which means the variety of frequencies in music isn’t represented well.''} CIs also did not bring perfect music experience for some, \textit{``I’m deaf since I was 2 and I also had CI ever since I was 3. Thanks to those 2 years I can speak normally and everything, and my CI’s are absolutely great which makes me able to listen to music almost the same way. I do have trouble knowing what instrument is playing and what notes and if those are high or low. I can actually understand some lyrics but never a lot of it.''} Another Reddit user who was not born deaf and had listened to music before they became deaf was frustrated by the current suboptimal music experience, \textit{``
Hearing and understanding music, or pretty much anything are 2 different things. If you mean hearing hearing, then yes, I do, if you mean understanding what I'm hearing, then no I don't. I hear the sound, I just can't make out any music out of it. Then again I wasn't born deaf or hearing impaired, so I remember what music sounded like, which is an added difficulty.''}

In response, DHH people often shared accessible songs with the community, \textit{``I tend to listen with strong vocal songs or using some sort of equalizer to make the vocal stronger. I tend to read the lyrics, seeing if I like it or not. Other times I listen just purely for the combination of the sounds.''} In addition to equalizers mentioned in the quote above, DHH people also actively shared other assistive technologies such as hearing aids and music devices like headphones. The sharing happened in either Q\&A posts or standalone posts for information sharing. For example, one user shared their special setup on the hearing aid programming that helped make music more sharp and crisp. Another user shared the combination of Apple earbuds and hearing aids. Specific hearing aids such as Phonak Audéo Paradise and Oticon S1 were also recommended. 
Big-DJ-type over-ear headphones, Bluetooth headphones with ambient sound settings, and specific brands and models such as Dome, 2E1, and sony mx5 were suggested.

\noindent \textbf{Medical Q\&A.} When one user asked if loud music could worsen deafness, several others provided their opinions and suggestions. One shared their experience of worsened hearing due to loud music, \textit{``I didn’t lose my hearing because of loud music but it’s 100\% gotten worse because of loud music. Hearing loss is also well known to cause hearing loss according to the Center of Disease Control. I’d find an alternative to the loud music sooner over later if that’s possible.''} Another person suggested protecting hearing by using noise-canceling headphones to remove background noise and listening to music at more normal levels. A user who was genetically HoH worried less about loud music, \textit{``I am still losing my hearing, but I get plenty of enjoyment from music now just as I did before. I have trouble understanding lyrics and high pitched guitar/piano/punchy drums don’t come through but that’s ok I just turn up the bass.''}

A HoH person just got their first hearing aid and experienced noises, preventing them from focusing. Other users provided suggestions based on their own experience, e.g., \textit{``The audiologist will likely tell you to use your hearing aid as much as you can, but you can definitely take it off when you're overwhelmed, or at least lower the volume.''} 
Possible reasons were also provided, \textit{``You’re hearing tons of stuff you haven’t heard in a long time, and your brain is going to take some time to adjust. Car noise, the turn signal, light switches, your feet, paper rustling. All of this is going to sound overwhelming. It will get better with time. The more you wear them, the more you’ll get used to it, and it won’t seem so overwhelming. Cars can be particularly bad. They are noisier than most resize.''} Someone shared a similar experience, \textit{``You're one day in, so this type of reaction is completely normal. I think it took me a solid 4 months until I could reliably wear my HAs all day long, and I still have days, over 3 years later, that I just have to take them off in some situations.''}

\section{Discussion}
Our analysis revealed diverse experiences and challenges from DHH communities' discussions around music. DHH people had diverse preferences about music and expressed various accessibility challenges. Assistive technologies developed in academia were not common in DHH people's discussions; they alternatively used their own heuristics to approach music, such as choosing specific music types, using sign language to interpret lyrics, and relying on haptic or visual cues. Information sharing was common in the DHH community regarding both music experiences and hearing loss. The insights drawn are crucial for developers of assistive technologies and for advocates who strive to cultivate more inclusive music environments. Addressing the challenges reflected in the negative sentiments and discourses around music while continuing to support and enhance the positive discourse that prevails within the DHH community is important. This balanced approach is essential for fostering an ecosystem that not only acknowledges and resolves difficulties but also celebrates and builds upon the positive experiences of the DHH community.

\subsection{Design Justice}
DHH people's music experience can be best understood in the context of deaf culture, guided by such questions as ``How does society perceive the deaf?'' ``How do they see themselves in society?'' and ``How do they identify themselves?'' \cite{de2017musical}. Both similarities and differences exist between the music experience of deaf people and hearing people -- according to Darrow, deaf people interact with music in ways similar to hearing people but less often engage in ritual uses of music \cite{darrow1993role}. 

Having a situated understanding of how DHH people listen to and interact with music is important, and possible. Music students' knowledge about deaf children's music experience increased after social, musical, and educational interactions with them -- they were able to better relate to the deaf population and showed interest in more knowledge acquisition \cite{kaiser2000effect}. Without an adequate understanding, hearing people tended to impose wrong and biased assumptions about DHH people's music experience, as evidenced in the Reddit discussions. For example, hearing people may assume that all DHH people listen to music by default or that every DHH individual has the same preference for music. The DHH community resisted such assumptions, stating that they had diverse choices and preferences about music and should not be treated as a ``monolith.'' In addition, deafness was commonly referred to as a spectrum, which led to different music experiences. DHH people’s preferences for music were also contextual -- even music lovers preferred not to have music when conversing with others to avoid interference. 

Understanding DHH people's diverse experiences and expectations of music can lead to designs that truly serve their needs. Several DHH people mentioned that they preferred to enjoy music alone; when they were included in group musical activities with other hearing people, they had to pretend to follow along -- leading to awkward and frustrating experiences. Thus they may not be willing to collaborate with hearing people to create short beats, as proposed in a previous study \cite{soderberg2016music}. There has only been a small body of literature on understanding DHH people's music experience. Ohshiro and Cartwright surveyed 50 DHH people to understand their use of technology and the barriers they faced in creative sound activities \cite{ohshiro2022people}. 
Their barriers to technology included unknown availability and limited options, echoing our results. Watkins presented a content analysis of online weblogs, vlogs, videos, and articles written by DHH people regarding music \cite{watkins2017deaf} -- most discussed topics were music experienced through visual/vibratory methods, music and Deaf Culture, personal fulfillment through music, the importance of music, and preference for musical instruments. We further provide a contextual understanding of DHH people's music discussions in Reddit deaf and HoH sub-communities, revealing both individualized experiences and community dynamics.

\subsection{Design Implications}
Hearing loss creates barriers to DHH people's optimal music experience. Drawing on the literature and our findings, we propose design implications to accommodate their challenges with music. 

\noindent \textbf{Building Accessibility Infrastructures.} Emerging technologies are often found inaccessible \cite{zhou2023iterative}. Deafness-related accessibility features are similarly lacking in many popular platforms and media. On Reddit, DHH people often complained about movies or online videos that did not provide captions for music. For example, they found difficulty receiving YouTube music videos and TikTok videos with background music if captioning was not provided. 
Several DHH users raved about the captioning function provided by some smartphones like iPhone and Google Pixel. According to them, Netflix was a good example of providing contextual captions for music in movies. Other streaming and video-based social platforms are encouraged to follow suit. AI-based music captioning technologies \cite{manco2021muscaps} may be of help. Music streaming platforms such as Spotify and Apple Music should make sure lyrics are always available for lyrical songs and consider accompanying music with ASL videos. 

\noindent \textbf{Tailored Music Recommendation and Generation.} DHH people have unique preferences compared to hearing people. For example, they tend to listen to music that they are familiar with. DHH people commonly express a preference for non-lyrical, instrument-heavy, and loud music, which is more accessible to them. Music streaming apps may consider providing customized playlists and music recommendations for DHH users. Further, generative AI models \cite{dong2018musegan} can be leveraged to generate music tailored to DHH people's needs, or allow DHH people to compose music for themselves.


\noindent \textbf{Democratizing Assistive Technologies.} Similar to \cite{watkins2017deaf}, our analysis revealed the importance of haptic and visual cues for enhancing DHH users' music experience. Haptic or visual-based assistive technologies were shown effective in research \cite{cavdir2020felt, jack2015designing, deja2020vitune, grierson2011making}, but have not been widely adopted by the DHH community \cite{ohshiro2022people}. The academic or industrial projects the Reddit users mentioned have not translated into easily purchasable products, and few in the DHH community mentioned using them. Affordability of assistive wearable devices should be given more consideration in the design process \cite{trivedi2019wearable}. Assistive technologies are important since DHH people's own heuristics did not always work -- playing music loud may bother family members or neighbors, especially in quiet hours -- and CIs and hearing aids did not always produce optimal music experience, according to them.

\noindent \textbf{Leveraging The DHH Community For Better Music Experience.} Marginalized communities are often found to nurture mutual support among their members \cite{tang2023towards, tang2023community}. In our case, DHH people tended to share accessible music, assistive technologies, and music devices with the community. They also frankly shared their suboptimal music experience and answered medical questions regarding hearing loss and hearing aids based on their own experience. Empathetic responses helped DHH people achieve a better music experience or adapt to hearing loss. Designers should actively refer to social media discussions to gauge DHH people's unique needs with music.


\section{Conclusion}
We presented a mixed-methods analysis of DHH people's online music discussions in their communities. They drew on their own heuristics for better music experience, such as ASL, visual/haptic cues, defaulting to familiar music, listening to non-lyrical, instrument-heavy music, and playing music loud. A reciprocal community was formed around music, where the members actively shared information and answered questions about music and hearing loss. Based on the findings, we discuss the design justice for enhancing DHH people's music experience through a more thorough understanding of their real, diverse musical needs. We conclude by providing concrete design implications. More research is encouraged to understand DHH people's music needs and how assistive music technologies help them in practice.




\bibliographystyle{ACM-Reference-Format}
\bibliography{sample-base}


\begin{thebibliography}{56}


\ifx \showCODEN    \undefined \def \showCODEN     #1{\unskip}     \fi
\ifx \showDOI      \undefined \def \showDOI       #1{#1}\fi
\ifx \showISBNx    \undefined \def \showISBNx     #1{\unskip}     \fi
\ifx \showISBNxiii \undefined \def \showISBNxiii  #1{\unskip}     \fi
\ifx \showISSN     \undefined \def \showISSN      #1{\unskip}     \fi
\ifx \showLCCN     \undefined \def \showLCCN      #1{\unskip}     \fi
\ifx \shownote     \undefined \def \shownote      #1{#1}          \fi
\ifx \showarticletitle \undefined \def \showarticletitle #1{#1}   \fi
\ifx \showURL      \undefined \def \showURL       {\relax}        \fi
\providecommand\bibfield[2]{#2}
\providecommand\bibinfo[2]{#2}
\providecommand\natexlab[1]{#1}
\providecommand\showeprint[2][]{arXiv:#2}

\bibitem[Aker et~al\mbox{.}(2022)]%
        {aker2022effect}
\bibfield{author}{\bibinfo{person}{Scott~C Aker}, \bibinfo{person}{Hamish Innes-Brown}, \bibinfo{person}{Kathleen~F Faulkner}, \bibinfo{person}{Marianna Vatti}, {and} \bibinfo{person}{Jeremy Marozeau}.} \bibinfo{year}{2022}\natexlab{}.
\newblock \showarticletitle{Effect of audio-tactile congruence on vibrotactile music enhancement}.
\newblock \bibinfo{journal}{\emph{The Journal of the Acoustical Society of America}} \bibinfo{volume}{152}, \bibinfo{number}{6} (\bibinfo{year}{2022}), \bibinfo{pages}{3396--3409}.
\newblock


\bibitem[Ara{\'u}jo and Batista(2016)]%
        {araujo2016auris}
\bibfield{author}{\bibinfo{person}{Felipe~Alves Ara{\'u}jo} {and} \bibinfo{person}{Carlos~Eduardo Batista}.} \bibinfo{year}{2016}\natexlab{}.
\newblock \showarticletitle{Auris: system for facilitating the musical perception for the hearing impaired}. In \bibinfo{booktitle}{\emph{Proceedings of the 22nd Brazilian Symposium on Multimedia and the Web}}. \bibinfo{pages}{135--142}.
\newblock


\bibitem[Baijal et~al\mbox{.}(2012)]%
        {baijal2012composing}
\bibfield{author}{\bibinfo{person}{Anant Baijal}, \bibinfo{person}{Julia Kim}, \bibinfo{person}{Carmen Branje}, \bibinfo{person}{Frank Russo}, {and} \bibinfo{person}{Deborah~I Fels}.} \bibinfo{year}{2012}\natexlab{}.
\newblock \showarticletitle{Composing vibrotactile music: A multi-sensory experience with the emoti-chair}. In \bibinfo{booktitle}{\emph{2012 ieee haptics symposium (haptics)}}. IEEE, \bibinfo{pages}{509--515}.
\newblock


\bibitem[Bassiouny et~al\mbox{.}(2017)]%
        {bassiouny2017using}
\bibfield{author}{\bibinfo{person}{Samia~E Bassiouny}, \bibinfo{person}{Marwa~M Saleh}, \bibinfo{person}{DA Elrefaie}, {and} \bibinfo{person}{Mary~S Girgis}.} \bibinfo{year}{2017}\natexlab{}.
\newblock \showarticletitle{Using music therapy in (re) habilitation of prelingual deaf cochlear implant children}.
\newblock \bibinfo{journal}{\emph{Biom J Sci Tech Res}} \bibinfo{volume}{1}, \bibinfo{number}{1} (\bibinfo{year}{2017}), \bibinfo{pages}{105--110}.
\newblock


\bibitem[Bossey(2020)]%
        {bossey2020accessibility}
\bibfield{author}{\bibinfo{person}{Adrian Bossey}.} \bibinfo{year}{2020}\natexlab{}.
\newblock \showarticletitle{Accessibility all areas? UK live music industry perceptions of current practice and Information and Communication Technology improvements to accessibility for music festival attendees who are deaf or disabled}.
\newblock \bibinfo{journal}{\emph{International Journal of Event and Festival Management}} \bibinfo{volume}{11}, \bibinfo{number}{1} (\bibinfo{year}{2020}), \bibinfo{pages}{6--25}.
\newblock


\bibitem[Braun and Clarke(2012)]%
        {braun2012thematic}
\bibfield{author}{\bibinfo{person}{Virginia Braun} {and} \bibinfo{person}{Victoria Clarke}.} \bibinfo{year}{2012}\natexlab{}.
\newblock \bibinfo{booktitle}{\emph{Thematic analysis.}}
\newblock \bibinfo{publisher}{American Psychological Association}.
\newblock


\bibitem[Cavdir and Wang(2020)]%
        {cavdir2020felt}
\bibfield{author}{\bibinfo{person}{Doga Cavdir} {and} \bibinfo{person}{Ge Wang}.} \bibinfo{year}{2020}\natexlab{}.
\newblock \showarticletitle{Felt sound: A shared musical experience for the deaf and hard of hearing}. In \bibinfo{booktitle}{\emph{Proceedings of the 20th international conference on new interfaces for musical expression (nime-20)}}.
\newblock


\bibitem[Chao et~al\mbox{.}(2018)]%
        {chao2018dancevibe}
\bibfield{author}{\bibinfo{person}{Chi-Ju Chao}, \bibinfo{person}{Chun-Wei Huang}, \bibinfo{person}{Chuan-Jie Lin}, \bibinfo{person}{Hao-Hua Chu}, {and} \bibinfo{person}{Polly Huang}.} \bibinfo{year}{2018}\natexlab{}.
\newblock \showarticletitle{DanceVibe: Assistive Dancing for the Hearing Impaired}. In \bibinfo{booktitle}{\emph{Mobile Computing, Applications, and Services: 9th International Conference, MobiCASE 2018, Osaka, Japan, February 28--March 2, 2018, Proceedings 9}}. Springer, \bibinfo{pages}{21--39}.
\newblock


\bibitem[Darrow(1993)]%
        {darrow1993role}
\bibfield{author}{\bibinfo{person}{Alice-Ann Darrow}.} \bibinfo{year}{1993}\natexlab{}.
\newblock \showarticletitle{The role of music in deaf culture: Implications for music educators}.
\newblock \bibinfo{journal}{\emph{Journal of Research in Music Education}} \bibinfo{volume}{41}, \bibinfo{number}{2} (\bibinfo{year}{1993}), \bibinfo{pages}{93--110}.
\newblock


\bibitem[De~Paula and Pederiva(2017)]%
        {de2017musical}
\bibfield{author}{\bibinfo{person}{Tatiane Ribeiro~Morais De~Paula} {and} \bibinfo{person}{Patr{\'\i}cia Lima~Martins Pederiva}.} \bibinfo{year}{2017}\natexlab{}.
\newblock \showarticletitle{Musical Experience in Deaf Culture}.
\newblock \bibinfo{journal}{\emph{International Journal of Technology and Inclusive Education}} \bibinfo{volume}{6}, \bibinfo{number}{2} (\bibinfo{year}{2017}), \bibinfo{pages}{1098--1107}.
\newblock


\bibitem[Deja et~al\mbox{.}(2020)]%
        {deja2020vitune}
\bibfield{author}{\bibinfo{person}{Jordan~Aiko Deja}, \bibinfo{person}{Alexczar Dela~Torre}, \bibinfo{person}{Hans~Joshua Lee}, \bibinfo{person}{Jose~Florencio Ciriaco~IV}, {and} \bibinfo{person}{Carlo~Miguel Eroles}.} \bibinfo{year}{2020}\natexlab{}.
\newblock \showarticletitle{Vitune: A visualizer tool to allow the deaf and hard of hearing to see music with their eyes}. In \bibinfo{booktitle}{\emph{Extended Abstracts of the 2020 CHI Conference on Human Factors in Computing Systems}}. \bibinfo{pages}{1--8}.
\newblock


\bibitem[Diyasa et~al\mbox{.}(2021)]%
        {diyasa2021twitter}
\bibfield{author}{\bibinfo{person}{I~Gede Susrama~Mas Diyasa}, \bibinfo{person}{Ni~Made Ika~Marini Mandenni}, \bibinfo{person}{Mohammad~Idham Fachrurrozi}, \bibinfo{person}{Sunu~Ilham Pradika}, \bibinfo{person}{Kholilul Rachman~Nur Manab}, {and} \bibinfo{person}{Nyoman~Rahadi Sasmita}.} \bibinfo{year}{2021}\natexlab{}.
\newblock \showarticletitle{Twitter Sentiment Analysis as an Evaluation and Service Base On Python Textblob}. In \bibinfo{booktitle}{\emph{IOP Conference Series: Materials Science and Engineering}}, Vol.~\bibinfo{volume}{1125}. IOP Publishing, \bibinfo{pages}{012034}.
\newblock


\bibitem[Dong et~al\mbox{.}(2018)]%
        {dong2018musegan}
\bibfield{author}{\bibinfo{person}{Hao-Wen Dong}, \bibinfo{person}{Wen-Yi Hsiao}, \bibinfo{person}{Li-Chia Yang}, {and} \bibinfo{person}{Yi-Hsuan Yang}.} \bibinfo{year}{2018}\natexlab{}.
\newblock \showarticletitle{Musegan: Multi-track sequential generative adversarial networks for symbolic music generation and accompaniment}. In \bibinfo{booktitle}{\emph{Proceedings of the AAAI Conference on Artificial Intelligence}}, Vol.~\bibinfo{volume}{32}.
\newblock


\bibitem[Elor et~al\mbox{.}(2021)]%
        {elor2021understanding}
\bibfield{author}{\bibinfo{person}{Aviv Elor}, \bibinfo{person}{Asiiah Song}, {and} \bibinfo{person}{Sri Kurniawan}.} \bibinfo{year}{2021}\natexlab{}.
\newblock \showarticletitle{Understanding emotional expression with haptic feedback vest patterns and immersive virtual reality}. In \bibinfo{booktitle}{\emph{2021 IEEE Conference on Virtual Reality and 3D User Interfaces Abstracts and Workshops (VRW)}}. IEEE, \bibinfo{pages}{183--188}.
\newblock


\bibitem[Fahey and Birkenshaw(1972)]%
        {fahey1972education}
\bibfield{author}{\bibinfo{person}{Joan~Dahms Fahey} {and} \bibinfo{person}{Lois Birkenshaw}.} \bibinfo{year}{1972}\natexlab{}.
\newblock \showarticletitle{Education of the deaf bypassing the ear: the perception of music by feeling and touch}.
\newblock \bibinfo{journal}{\emph{Music Educators Journal}} \bibinfo{volume}{58}, \bibinfo{number}{8} (\bibinfo{year}{1972}), \bibinfo{pages}{44--51}.
\newblock


\bibitem[Florian et~al\mbox{.}(2017)]%
        {florian2017deaf}
\bibfield{author}{\bibinfo{person}{Horațiu Florian}, \bibinfo{person}{Adrian Mocanu}, \bibinfo{person}{Cristian Vlasin}, \bibinfo{person}{Jos{\'e} Machado}, \bibinfo{person}{Vitor Carvalho}, \bibinfo{person}{Filomena Soares}, \bibinfo{person}{Adina Astilean}, {and} \bibinfo{person}{Camelia Avram}.} \bibinfo{year}{2017}\natexlab{}.
\newblock \showarticletitle{Deaf people feeling music rhythm by using a sensing and actuating device}.
\newblock \bibinfo{journal}{\emph{Sensors and Actuators A: Physical}}  \bibinfo{volume}{267} (\bibinfo{year}{2017}), \bibinfo{pages}{431--442}.
\newblock


\bibitem[Fourney and Fels(2009)]%
        {fourney2009creating}
\bibfield{author}{\bibinfo{person}{David~W Fourney} {and} \bibinfo{person}{Deborah~I Fels}.} \bibinfo{year}{2009}\natexlab{}.
\newblock \showarticletitle{Creating access to music through visualization}. In \bibinfo{booktitle}{\emph{2009 ieee toronto international conference science and technology for humanity (tic-sth)}}. IEEE, \bibinfo{pages}{939--944}.
\newblock


\bibitem[Frid(2019)]%
        {frid2019accessible}
\bibfield{author}{\bibinfo{person}{Emma Frid}.} \bibinfo{year}{2019}\natexlab{}.
\newblock \showarticletitle{Accessible digital musical instruments—a review of musical interfaces in inclusive music practice}.
\newblock \bibinfo{journal}{\emph{Multimodal Technologies and Interaction}} \bibinfo{volume}{3}, \bibinfo{number}{3} (\bibinfo{year}{2019}), \bibinfo{pages}{57}.
\newblock


\bibitem[Fulford et~al\mbox{.}(2015)]%
        {fulford2015hearing}
\bibfield{author}{\bibinfo{person}{Robert Fulford}, \bibinfo{person}{Jane Ginsborg}, {and} \bibinfo{person}{Alinka Greasley}.} \bibinfo{year}{2015}\natexlab{}.
\newblock \showarticletitle{Hearing aids and music: the experiences of D/deaf musicians}. In \bibinfo{booktitle}{\emph{Ninth Triennial Conference of the European Society for the Cognitive Sciences of Music, Manchester, UK, August}}. \bibinfo{pages}{17--22}.
\newblock


\bibitem[Grierson(2011)]%
        {grierson2011making}
\bibfield{author}{\bibinfo{person}{Michael~S Grierson}.} \bibinfo{year}{2011}\natexlab{}.
\newblock \showarticletitle{Making music with images: interactive audiovisual performance systems for the deaf}.
\newblock  (\bibinfo{year}{2011}).
\newblock


\bibitem[Han and Kim(2023)]%
        {han2023opportunities}
\bibfield{author}{\bibinfo{person}{Hyomin Han} {and} \bibinfo{person}{Taewook Kim}.} \bibinfo{year}{2023}\natexlab{}.
\newblock \showarticletitle{Opportunities to Support Communal Experiences of Deaf and Hard-of-Hearing People in Live Popular Music Concerts}. In \bibinfo{booktitle}{\emph{Companion Proceedings of the 2023 ACM International Conference on Supporting Group Work}}. \bibinfo{pages}{36--38}.
\newblock


\bibitem[Hopyan et~al\mbox{.}(2011)]%
        {hopyan2011identifying}
\bibfield{author}{\bibinfo{person}{T Hopyan}, \bibinfo{person}{KA Gordon}, {and} \bibinfo{person}{BC Papsin}.} \bibinfo{year}{2011}\natexlab{}.
\newblock \showarticletitle{Identifying emotions in music through electrical hearing in deaf children using cochlear implants}.
\newblock \bibinfo{journal}{\emph{Cochlear implants international}} \bibinfo{volume}{12}, \bibinfo{number}{1} (\bibinfo{year}{2011}), \bibinfo{pages}{21--26}.
\newblock


\bibitem[Iijima et~al\mbox{.}(2022)]%
        {iijima2022designing}
\bibfield{author}{\bibinfo{person}{Ryo Iijima}, \bibinfo{person}{Akihisa Shitara}, {and} \bibinfo{person}{Yoichi Ochiai}.} \bibinfo{year}{2022}\natexlab{}.
\newblock \showarticletitle{Designing Gestures for Digital Musical Instruments: Gesture Elicitation Study with Deaf and Hard of Hearing People}. In \bibinfo{booktitle}{\emph{Proceedings of the 24th International ACM SIGACCESS Conference on Computers and Accessibility}}. \bibinfo{pages}{1--8}.
\newblock


\bibitem[Iijima et~al\mbox{.}(2021)]%
        {iijima2021smartphone}
\bibfield{author}{\bibinfo{person}{Ryo Iijima}, \bibinfo{person}{Akihisa Shitara}, \bibinfo{person}{Sayan Sarcar}, {and} \bibinfo{person}{Yoichi Ochiai}.} \bibinfo{year}{2021}\natexlab{}.
\newblock \showarticletitle{Smartphone Drum: Gesture-based Digital Musical Instruments Application for Deaf and Hard of Hearing People}. In \bibinfo{booktitle}{\emph{Proceedings of the 2021 ACM Symposium on Spatial User Interaction}}. \bibinfo{pages}{1--2}.
\newblock


\bibitem[JACK et~al\mbox{.}({[n.\,d.]})]%
        {jackdesign}
\bibfield{author}{\bibinfo{person}{ROBERT JACK}, \bibinfo{person}{ANDREW MCPHERSON}, {and} \bibinfo{person}{TONY STOCKMAN}.} \bibinfo{year}{[n.\,d.]}\natexlab{}.
\newblock \showarticletitle{The design of tactile musical devices for the deaf}.
\newblock  (\bibinfo{year}{[n.\,d.]}).
\newblock


\bibitem[Jack et~al\mbox{.}(2015)]%
        {jack2015designing}
\bibfield{author}{\bibinfo{person}{Robert Jack}, \bibinfo{person}{Andrew McPherson}, {and} \bibinfo{person}{Tony Stockman}.} \bibinfo{year}{2015}\natexlab{}.
\newblock \showarticletitle{Designing tactile musical devices with and for deaf users: a case study}. In \bibinfo{booktitle}{\emph{Proceedings of the International Conference on the Multimodal Experience of Music, Sheffield, UK}}. \bibinfo{pages}{23--25}.
\newblock


\bibitem[Kaiser and Johnson(2000)]%
        {kaiser2000effect}
\bibfield{author}{\bibinfo{person}{Keith~A Kaiser} {and} \bibinfo{person}{Krista~E Johnson}.} \bibinfo{year}{2000}\natexlab{}.
\newblock \showarticletitle{The effect of an interactive experience on music majors' perceptions of music for deaf students}.
\newblock \bibinfo{journal}{\emph{Journal of Music Therapy}} \bibinfo{volume}{37}, \bibinfo{number}{3} (\bibinfo{year}{2000}), \bibinfo{pages}{222--234}.
\newblock


\bibitem[Karam et~al\mbox{.}(2009)]%
        {karam2009modelling}
\bibfield{author}{\bibinfo{person}{Maria Karam}, \bibinfo{person}{Gabe Nespoli}, \bibinfo{person}{Frank Russo}, {and} \bibinfo{person}{Deborah~I Fels}.} \bibinfo{year}{2009}\natexlab{}.
\newblock \showarticletitle{Modelling perceptual elements of music in a vibrotactile display for deaf users: A field study}. In \bibinfo{booktitle}{\emph{2009 Second International Conferences on Advances in Computer-Human Interactions}}. IEEE, \bibinfo{pages}{249--254}.
\newblock


\bibitem[Kim et~al\mbox{.}(2015)]%
        {kim2015seen}
\bibfield{author}{\bibinfo{person}{Jeeeun Kim}, \bibinfo{person}{Swamy Ananthanarayan}, {and} \bibinfo{person}{Tom Yeh}.} \bibinfo{year}{2015}\natexlab{}.
\newblock \showarticletitle{Seen music: ambient music data visualization for children with hearing impairments}. In \bibinfo{booktitle}{\emph{Proceedings of the 14th International Conference on Interaction Design and Children}}. \bibinfo{pages}{426--429}.
\newblock


\bibitem[Kyriakou(2022)]%
        {kyriakou2022teaching}
\bibfield{author}{\bibinfo{person}{Karen Kyriakou}.} \bibinfo{year}{2022}\natexlab{}.
\newblock \showarticletitle{Teaching music to deaf students: A personal reflection}.
\newblock \bibinfo{journal}{\emph{Australian Journal of Music Education}} \bibinfo{volume}{54}, \bibinfo{number}{2} (\bibinfo{year}{2022}), \bibinfo{pages}{52--59}.
\newblock


\bibitem[Maler(2015)]%
        {maler2015musical}
\bibfield{author}{\bibinfo{person}{Anabel Maler}.} \bibinfo{year}{2015}\natexlab{}.
\newblock \showarticletitle{Musical expression among deaf and hearing song signers}.
\newblock \bibinfo{journal}{\emph{The Oxford handbook of music and disability studies}} (\bibinfo{year}{2015}), \bibinfo{pages}{73--91}.
\newblock


\bibitem[Manco et~al\mbox{.}(2021)]%
        {manco2021muscaps}
\bibfield{author}{\bibinfo{person}{Ilaria Manco}, \bibinfo{person}{Emmanouil Benetos}, \bibinfo{person}{Elio Quinton}, {and} \bibinfo{person}{Gy{\"o}rgy Fazekas}.} \bibinfo{year}{2021}\natexlab{}.
\newblock \showarticletitle{Muscaps: Generating captions for music audio}. In \bibinfo{booktitle}{\emph{2021 International Joint Conference on Neural Networks (IJCNN)}}. IEEE, \bibinfo{pages}{1--8}.
\newblock


\bibitem[McHugh et~al\mbox{.}(2021)]%
        {mchugh2021towards}
\bibfield{author}{\bibinfo{person}{Thomas~Barlow McHugh}, \bibinfo{person}{Abir Saha}, \bibinfo{person}{David Bar-El}, \bibinfo{person}{Marcelo Worsley}, {and} \bibinfo{person}{Anne~Marie Piper}.} \bibinfo{year}{2021}\natexlab{}.
\newblock \showarticletitle{Towards Inclusive Streaming: Building Multimodal Music Experiences for the Deaf and Hard of Hearing}. In \bibinfo{booktitle}{\emph{Extended Abstracts of the 2021 CHI Conference on Human Factors in Computing Systems}}. \bibinfo{pages}{1--7}.
\newblock


\bibitem[Nanayakkara et~al\mbox{.}(2009)]%
        {nanayakkara2009enhanced}
\bibfield{author}{\bibinfo{person}{Suranga Nanayakkara}, \bibinfo{person}{Elizabeth Taylor}, \bibinfo{person}{Lonce Wyse}, {and} \bibinfo{person}{S~H Ong}.} \bibinfo{year}{2009}\natexlab{}.
\newblock \showarticletitle{An enhanced musical experience for the deaf: design and evaluation of a music display and a haptic chair}. In \bibinfo{booktitle}{\emph{Proceedings of the sigchi conference on human factors in computing systems}}. \bibinfo{pages}{337--346}.
\newblock


\bibitem[Nanayakkara et~al\mbox{.}(2013)]%
        {nanayakkara2013enhancing}
\bibfield{author}{\bibinfo{person}{Suranga~Chandima Nanayakkara}, \bibinfo{person}{Lonce Wyse}, \bibinfo{person}{Sim~Heng Ong}, {and} \bibinfo{person}{Elizabeth~A Taylor}.} \bibinfo{year}{2013}\natexlab{}.
\newblock \showarticletitle{Enhancing musical experience for the hearing-impaired using visual and haptic displays}.
\newblock \bibinfo{journal}{\emph{Human--Computer Interaction}} \bibinfo{volume}{28}, \bibinfo{number}{2} (\bibinfo{year}{2013}), \bibinfo{pages}{115--160}.
\newblock


\bibitem[Ohshiro and Cartwright(2022)]%
        {ohshiro2022people}
\bibfield{author}{\bibinfo{person}{Keita Ohshiro} {and} \bibinfo{person}{Mark Cartwright}.} \bibinfo{year}{2022}\natexlab{}.
\newblock \showarticletitle{How people who are deaf, Deaf, and hard of hearing use technology in creative sound activities}. In \bibinfo{booktitle}{\emph{Proceedings of the 24th International ACM SIGACCESS Conference on Computers and Accessibility}}. \bibinfo{pages}{1--4}.
\newblock


\bibitem[Palmer and Ojala(2022)]%
        {palmer2022vibrational}
\bibfield{author}{\bibinfo{person}{Russ~C Palmer} {and} \bibinfo{person}{Stina Ojala}.} \bibinfo{year}{2022}\natexlab{}.
\newblock \showarticletitle{Vibrational Music Therapy with D/deaf clients}. In \bibinfo{booktitle}{\emph{Voices: A World Forum for Music Therapy}}, Vol.~\bibinfo{volume}{22}.
\newblock


\bibitem[Petry et~al\mbox{.}(2018)]%
        {petry2018supporting}
\bibfield{author}{\bibinfo{person}{Benjamin Petry}, \bibinfo{person}{Thavishi Illandara}, \bibinfo{person}{Don~Samitha Elvitigala}, {and} \bibinfo{person}{Suranga Nanayakkara}.} \bibinfo{year}{2018}\natexlab{}.
\newblock \showarticletitle{Supporting rhythm activities of deaf children using music-sensory-substitution systems}. In \bibinfo{booktitle}{\emph{Proceedings of the 2018 CHI Conference on Human Factors in Computing Systems}}. \bibinfo{pages}{1--10}.
\newblock


\bibitem[Petry et~al\mbox{.}(2016b)]%
        {petry2016ad}
\bibfield{author}{\bibinfo{person}{Benjamin Petry}, \bibinfo{person}{Thavishi Illandara}, \bibinfo{person}{Juan~Pablo Forero}, {and} \bibinfo{person}{Suranga Nanayakkara}.} \bibinfo{year}{2016}\natexlab{b}.
\newblock \showarticletitle{Ad-hoc access to musical sound for deaf individuals}. In \bibinfo{booktitle}{\emph{Proceedings of the 18th International ACM SIGACCESS Conference on Computers and Accessibility}}. \bibinfo{pages}{285--286}.
\newblock


\bibitem[Petry et~al\mbox{.}(2016a)]%
        {petry2016muss}
\bibfield{author}{\bibinfo{person}{Benjamin Petry}, \bibinfo{person}{Thavishi Illandara}, {and} \bibinfo{person}{Suranga Nanayakkara}.} \bibinfo{year}{2016}\natexlab{a}.
\newblock \showarticletitle{MuSS-bits: sensor-display blocks for deaf people to explore musical sounds}. In \bibinfo{booktitle}{\emph{Proceedings of the 28th Australian Conference on Computer-Human Interaction}}. \bibinfo{pages}{72--80}.
\newblock


\bibitem[Remache-Vinueza et~al\mbox{.}(2021)]%
        {remache2021audio}
\bibfield{author}{\bibinfo{person}{Byron Remache-Vinueza}, \bibinfo{person}{Andr{\'e}s Trujillo-Le{\'o}n}, \bibinfo{person}{Mireya Zapata}, \bibinfo{person}{Fabi{\'a}n Sarmiento-Ortiz}, {and} \bibinfo{person}{Fernando Vidal-Verd{\'u}}.} \bibinfo{year}{2021}\natexlab{}.
\newblock \showarticletitle{Audio-tactile rendering: a review on technology and methods to convey musical information through the sense of touch}.
\newblock \bibinfo{journal}{\emph{Sensors}} \bibinfo{volume}{21}, \bibinfo{number}{19} (\bibinfo{year}{2021}), \bibinfo{pages}{6575}.
\newblock


\bibitem[Rochette et~al\mbox{.}(2014)]%
        {rochette2014music}
\bibfield{author}{\bibinfo{person}{Fran{\c{c}}oise Rochette}, \bibinfo{person}{Aline Moussard}, {and} \bibinfo{person}{Emmanuel Bigand}.} \bibinfo{year}{2014}\natexlab{}.
\newblock \showarticletitle{Music lessons improve auditory perceptual and cognitive performance in deaf children}.
\newblock \bibinfo{journal}{\emph{Frontiers in human neuroscience}}  \bibinfo{volume}{8} (\bibinfo{year}{2014}), \bibinfo{pages}{488}.
\newblock


\bibitem[Schellenberg(2005)]%
        {schellenberg2005music}
\bibfield{author}{\bibinfo{person}{E~Glenn Schellenberg}.} \bibinfo{year}{2005}\natexlab{}.
\newblock \showarticletitle{Music and cognitive abilities}.
\newblock \bibinfo{journal}{\emph{Current Directions in Psychological Science}} \bibinfo{volume}{14}, \bibinfo{number}{6} (\bibinfo{year}{2005}), \bibinfo{pages}{317--320}.
\newblock


\bibitem[Silva et~al\mbox{.}(2020)]%
        {silva2020music}
\bibfield{author}{\bibinfo{person}{Nedinaldo Manoel~da Silva}, \bibinfo{person}{Jefferson~Fernandes Alves}, \bibinfo{person}{Ahiram Brunni Cartaxo~de Castro}, {and} \bibinfo{person}{Jed{\'\i}dja Hadassa de~Santana Varela}.} \bibinfo{year}{2020}\natexlab{}.
\newblock \showarticletitle{Music education for the deaf: characteristics, barriers and successful practices}.
\newblock \bibinfo{journal}{\emph{Educa{\c{c}}{\~a}o e Pesquisa}}  \bibinfo{volume}{46} (\bibinfo{year}{2020}).
\newblock


\bibitem[Sion et~al\mbox{.}(2023)]%
        {sion2023me}
\bibfield{author}{\bibinfo{person}{Yulia Sion}, \bibinfo{person}{Sunil Sudevan}, {and} \bibinfo{person}{David Lamas}.} \bibinfo{year}{2023}\natexlab{}.
\newblock \showarticletitle{Be Me Vest-Exploring the Emotional Effects of Music and Sound-Based Vibrotactile Stimuli}. In \bibinfo{booktitle}{\emph{International Conference on Human-Computer Interaction}}. Springer, \bibinfo{pages}{318--331}.
\newblock


\bibitem[S{\o}derberg et~al\mbox{.}(2016)]%
        {soderberg2016music}
\bibfield{author}{\bibinfo{person}{Ene~Alicia S{\o}derberg}, \bibinfo{person}{Rasmus~Emil Odgaard}, \bibinfo{person}{Sarah Bitsch}, \bibinfo{person}{Oliver H{\o}eg-Jensen}, \bibinfo{person}{Nikolaj~Schildt Christensen}, \bibinfo{person}{S{\o}ren~Dahl Poulsen}, {and} \bibinfo{person}{Steven Gelineck}.} \bibinfo{year}{2016}\natexlab{}.
\newblock \showarticletitle{Music Aid: Towards a Collaborative Experience for Deaf and Hearing People in Creating Music}. In \bibinfo{booktitle}{\emph{New Interfaces for Musical Expression}}.
\newblock


\bibitem[Tang et~al\mbox{.}(2023a)]%
        {tang2023community}
\bibfield{author}{\bibinfo{person}{Xinru Tang}, \bibinfo{person}{Xiang Chang}, \bibinfo{person}{Nuoran Chen}, \bibinfo{person}{Yingjie Ni}, \bibinfo{person}{RAY LC}, {and} \bibinfo{person}{Xin Tong}.} \bibinfo{year}{2023}\natexlab{a}.
\newblock \showarticletitle{Community-Driven Information Accessibility: Online Sign Language Content Creation within d/Deaf Communities}. In \bibinfo{booktitle}{\emph{Proceedings of the 2023 CHI Conference on Human Factors in Computing Systems}}. \bibinfo{pages}{1--24}.
\newblock


\bibitem[Tang et~al\mbox{.}(2023b)]%
        {tang2023towards}
\bibfield{author}{\bibinfo{person}{Xinru Tang}, \bibinfo{person}{Xianghua Ding}, {and} \bibinfo{person}{Zhixuan Zhou}.} \bibinfo{year}{2023}\natexlab{b}.
\newblock \showarticletitle{Towards Equitable Online Participation: A Case of Older Adult Content Creators' Role Transition on Short-form Video Sharing Platforms}.
\newblock \bibinfo{journal}{\emph{Proceedings of the ACM on Human-Computer Interaction}} \bibinfo{volume}{7}, \bibinfo{number}{CSCW2} (\bibinfo{year}{2023}), \bibinfo{pages}{1--22}.
\newblock


\bibitem[Tranchant et~al\mbox{.}(2017)]%
        {tranchant2017feeling}
\bibfield{author}{\bibinfo{person}{Pauline Tranchant}, \bibinfo{person}{Martha~M Shiell}, \bibinfo{person}{Marcello Giordano}, \bibinfo{person}{Alexis Nadeau}, \bibinfo{person}{Isabelle Peretz}, {and} \bibinfo{person}{Robert~J Zatorre}.} \bibinfo{year}{2017}\natexlab{}.
\newblock \showarticletitle{Feeling the beat: Bouncing synchronization to vibrotactile music in hearing and early deaf people}.
\newblock \bibinfo{journal}{\emph{Frontiers in neuroscience}}  \bibinfo{volume}{11} (\bibinfo{year}{2017}), \bibinfo{pages}{507}.
\newblock


\bibitem[Trivedi et~al\mbox{.}(2019)]%
        {trivedi2019wearable}
\bibfield{author}{\bibinfo{person}{Urvish Trivedi}, \bibinfo{person}{Redwan Alqasemi}, {and} \bibinfo{person}{Rajiv Dubey}.} \bibinfo{year}{2019}\natexlab{}.
\newblock \showarticletitle{Wearable musical haptic sleeves for people with hearing impairment}. In \bibinfo{booktitle}{\emph{Proceedings of the 12th ACM International Conference on Pervasive Technologies Related to Assistive Environments}}. \bibinfo{pages}{146--151}.
\newblock


\bibitem[Watkins(2017)]%
        {watkins2017deaf}
\bibfield{author}{\bibinfo{person}{Corinne~Scalia Watkins}.} \bibinfo{year}{2017}\natexlab{}.
\newblock \emph{\bibinfo{title}{The Deaf Perspective: A Content Analysis Study to Determine Deaf and Hard-of-Hearing Individuals' Perceptions and Attitudes Towards Music}}.
\newblock \bibinfo{thesistype}{Ph.\,D. Dissertation}. \bibinfo{school}{The Florida State University}.
\newblock


\bibitem[Wright(2014)]%
        {wright2014bridging}
\bibfield{author}{\bibinfo{person}{Whitney Wright}.} \bibinfo{year}{2014}\natexlab{}.
\newblock \showarticletitle{Bridging Music and the Early Childhood Curriculum in Listening and Spoken Language Programs for Children who are Deaf or Hard of Hearing}.
\newblock  (\bibinfo{year}{2014}).
\newblock


\bibitem[Yan and Liu(2022)]%
        {yan2022adaptive}
\bibfield{author}{\bibinfo{person}{Tianqiang Yan} {and} \bibinfo{person}{Yuwei Liu}.} \bibinfo{year}{2022}\natexlab{}.
\newblock \showarticletitle{An Adaptive Musical Vibrotactile System (MuViT) for Modern Smartphones}. In \bibinfo{booktitle}{\emph{International Forum on Digital TV and Wireless Multimedia Communications}}. Springer, \bibinfo{pages}{436--448}.
\newblock


\bibitem[Yoo et~al\mbox{.}(2023)]%
        {yoo2023understanding}
\bibfield{author}{\bibinfo{person}{Suhyeon Yoo}, \bibinfo{person}{Georgianna Lin}, \bibinfo{person}{Hyeon~Jeong Byeon}, \bibinfo{person}{Amy~S Hwang}, {and} \bibinfo{person}{Khai~Nhut Truong}.} \bibinfo{year}{2023}\natexlab{}.
\newblock \showarticletitle{Understanding tensions in music accessibility through song signing for and with d/Deaf and Non-d/Deaf persons}. In \bibinfo{booktitle}{\emph{Proceedings of the 2023 CHI Conference on Human Factors in Computing Systems}}. \bibinfo{pages}{1--18}.
\newblock


\bibitem[Young et~al\mbox{.}(2023)]%
        {young2023feel}
\bibfield{author}{\bibinfo{person}{Gareth~W Young}, \bibinfo{person}{N{\'e}ill O’Dwyer}, \bibinfo{person}{Mauricio~Flores Vargas}, \bibinfo{person}{Rachel~Mc Donnell}, {and} \bibinfo{person}{Aljosa Smolic}.} \bibinfo{year}{2023}\natexlab{}.
\newblock \showarticletitle{Feel the Music!—Audience Experiences of Audio--Tactile Feedback in a Novel Virtual Reality Volumetric Music Video}. In \bibinfo{booktitle}{\emph{Arts}}, Vol.~\bibinfo{volume}{12}. MDPI, \bibinfo{pages}{156}.
\newblock


\bibitem[Zhou et~al\mbox{.}(2023)]%
        {zhou2023iterative}
\bibfield{author}{\bibinfo{person}{Zhixuan Zhou}, \bibinfo{person}{Tanusree Sharma}, \bibinfo{person}{Luke Emano}, \bibinfo{person}{Sauvik Das}, {and} \bibinfo{person}{Yang Wang}.} \bibinfo{year}{2023}\natexlab{}.
\newblock \showarticletitle{Iterative Design of An Accessible Crypto Wallet for Blind Users}.
\newblock \bibinfo{journal}{\emph{arXiv preprint arXiv:2306.06261}} (\bibinfo{year}{2023}).
\newblock


\end{thebibliography}




\end{document}